# Signatures of two gaps in the spin susceptibility of a cuprate superconductor

Rui Zhou[1a], Igor Vinograd[1b], Michihiro Hirata[1c], Tao Wu[1d], Hadrien Mayaffre[1], Steffen Krämer[1], W.N. Hardy[2,3], Ruixing Liang[2,3], D.A. Bonn[2,3], Toshinao Loew[4], Juan Porras[4], Bernhard Keimer[4] and Marc-Henri Julien[1*]

[1]Laboratoire National des Champs Magnétiques Intenses, CNRS – Université Grenoble Alpes – Université Paul Sabatier – Institut National des Sciences Appliquées – European Magnetic Field Laboratory, 38042 Grenoble, France.

[2]Department of Physics and Astronomy, University of British Columbia, Vancouver, BC V6T 1Z1, Canada.

[3]Canadian Institute for Advanced Research, Toronto, Ontario M5G 1Z8, Canada.

[4]Max-Planck-Institut für Festkörperforschung, Heisenbergstrasse 1, D-70569 Stuttgart, Germany

[*]Corresponding author. Email: marc-henri.julien@lncmi.cnrs.fr

**A major obstacle to understanding high-$T_c$ cuprates is that superconductivity precludes observing normal-state properties at low temperatures. One prime example is the normal-state spin susceptibility $\chi_{spin}$: although its decrease upon cooling far above $T_c$ typifies pseudogap behavior, its behavior at low temperatures is generally unknown. Here, our measurements in high magnetic fields expose $\chi_{spin}$ of $YBa_2Cu_3O_y$ down to low temperatures. Even though superconductivity is suppressed by the field, we uncover two thermally-activated contributions alongside a residual $\chi_{spin}(T=0)$ due to gapless excitations. We relate these two distinct gaps to short-range charge-density waves and to the formation of singlets as in certain quantum spin systems. Both phenomena thus contribute to the pseudogap at low temperature, supplementing short-lived antiferromagnetism that initiates pseudogap behavior at high temperatures. We therefore propose that the pseudogap ought to be regarded as a composite property and that, when not undergoing spin-stripe ordering, underdoped cuprates tend to form short-ranged spin singlets.**

The pseudogap of the cuprate superconductors has fascinated researchers for more than three decades, not only because it is the most puzzling of their normal-state properties but also because it is considered to be the key to understanding the pairing mechanism. In recent years, discussions about the pseudogap have revolved mostly around single-particle properties or broken symmetries (1). Historically, however, the first use of the word 'pseudogap' in the cuprate context (2) was motivated by the observation that the uniform, static spin susceptibility $\chi_{spin} = \chi'_{spin}(q=0,\omega=0)$ decreases upon cooling in the metallic regime (2,3), as depicted in Fig. 1a.

In principle, there is every reason for $\chi_{spin}(T)$ to decrease in cuprates (4,5): antiferromagnetic (AFM) correlations are known to be ubiquitous and their growth upon cooling implies that magnetic spectral weight is transferred from the entire Brillouin zone (thus including $q$=0) to the AFM wave vector, which results in a decrease of $\chi_{spin}(q=0)$. This is most clearly observed in all low-dimensional antiferromagnets, owing to the large temperature range between the onset of AFM correlations (on the order of the main exchange interaction $J$) and the transition towards three-dimensional order, if any (6). A major question, however, is whether AFM correlations are the only explanation for the pseudogap in $\chi_{spin}$. Indeed, different proposals for the pseudogap phase, such as the formation of charge-density waves (CDW), of spin singlets or of incoherent Cooper pairs, are all susceptible to further reduce $\chi_{spin}$ beyond the effect of mere AFM correlations (Fig. 1).

Unraveling the pseudogap physics encoded in $\chi_{spin}(T)$ has, however, proved challenging: the ability to deconstruct $\chi_{spin}$ by modeling the data is severely limited by the occurrence of singlet-type superconductivity that masks the normal-state $\chi_{spin}$ from $T = 0$ up to temperatures on the order of $T_c \sim 100$ K. Given the current inability of state-of-the-art numerical techniques to calculate $\chi_{spin}(T)$ in this low $T$ range (5), the only viable solution to tackle this problem is to suppress superconductivity with intense magnetic fields ($B$) and to measure the normal state $\chi_{spin}$ down to the lowest temperature (7).

$YBa_2Cu_3O_y$ (YBCO) provides a unique opportunity to perform such measurements in a setting that combines minimal electronic inhomogeneity (8) with an experimentally accessible high-field regime over which $\chi_{spin}$ is affected neither by superconductivity (*i.e.* $\chi_{spin}$ is $B$ independent, ref. 9) nor by spin order (10). Furthermore, $\chi_{spin}$ shows no noticeable change upon crossing the field-induced transition towards long-range three-dimensional (3D) CDW order (10,11), as a function of either $B$ (9) or $T$ (12). This indicates that $\chi_{spin}$ is insensitive to whether the CDW is long- or short-ranged ordered, presumably because for the latter case the correlation length at $T \sim T_{CDW}$ is already substantial and larger than the magnetic one. Therefore, as far as $\chi_{spin}$ is concerned, the high-field data can be considered to faithfully reflect the $B = 0$ ground state without superconductivity.

To shed light on the normal-state $\chi_{spin}$, we have applied intense magnetic fields to suppress superconductivity in underdoped $YBa_2Cu_3O_y$ and we have used $^{17}$O and $^{89}$Y nuclear magnetic resonance (NMR) to measure the spin part of its Knight shift ($K_{spin}$), which is proportional to $\chi_{spin}$ of the $CuO_2$ planes:

$$K_{spin} = g\, \mu_B\, A_{hf}\, \chi_{spin} \qquad \text{(Eq. 1)}$$

where $A_{hf}$ is the hyperfine coupling, $g$ the Landé factor. Unlike bulk measurements, the Knight shift is not contaminated by the chain $\chi_{spin}$.

The bulk of the data is shown in Fig. 2 (the construction of this dataset is detailed in Methods). At the qualitative level, all samples show a similar trend: $K_{spin}$ continuously drops upon cooling and reaches a finite value $K_{res}$ as $T \to 0$. At the quantitative level, however, there are significant differences as a function of doping level $p$, that we now proceed to analyze.

**The charge-ordered regime**

We first discuss results in the doping range $p$=0.072–0.135. Even though the superconducting gap has been quenched in high fields, the low temperature behavior turns out to be accurately described by an activated form (plots of the data in logarithmic scale show that the decay is indeed activated and not *e.g*. a power law, see Supplementary Fig. S1):

$K_{spin}(T) = K_{res} + K_L \exp(-\Delta_L/k_B T)$. (Eq. 2)

Other variants of exponential decays also fit the data and were all found to yield consistent results (see SM and Figs. S2 and S3). Fit results based on simple exponentials (Eqs. 2,3,4) and on a hyperbolic tangent function are shown separately in Fig. S2. Since there is no strong reason to favor one function over the other, the fit results shown in Fig. 3 are the mean values of the numbers obtained by the two procedures. In the text, we keep the simple exponential function for the sake of simplicity.

These findings show that, in the absence of superconductivity, there is a finite, residual $\chi_{spin}(T{=}0) \propto K_{res}$ (to be discussed later) but also a well-defined gap $\Delta_L$ in the spin excitations at $q$=0. Quite surprisingly however, the obtained $\Delta_L$ values decrease upon decreasing $p$, which is opposite to the pseudogap energy scale (1,38). Inspection of the phase diagram (Fig. 3a) reveals that $\Delta_L$ actually shows the same doping dependence as $T_{CDW}$, the onset temperature of 2D CDW correlations, extrapolating to zero at $p \approx 0.06$–$0.07$ (13,14,15). In other words, the gap value turns out to be proportional to the characteristic temperature of CDW formation, namely $\Delta_L/k_B = \alpha\ T_{CDW}$ with $\alpha$ of order one (1–1.3 depending on the fit function, see Methods and Fig. S3). Therefore, in the absence of superconductivity, the low temperature $\chi_{spin}$ shows thermally activated behavior related to the CDW.

Connections between (certain aspects of) the pseudogap and the short-range CDW have been unveiled in earlier studies (16-22) but our findings offer a complementary perspective: while they obviously fall short of attributing the overall responsibility for the pseudogap solely to the CDW, they imply that, when present, the CDW state contributes to the pseudogap in $\chi_{spin}$.

How the CDW affects $\chi_{spin}$ can be conceptualized in essentially two different ways. First, the opening of the CDW gap around the Fermi energy $E_F$ can make $\chi_{spin}$ to drop exponentially below $T_{CDW}$, as observed in a number of 2D metals (Fig. 1e and refs. 23,24). Alternatively, the CDW may be intertwined with a form of spin pairing. Theoretical proposals along this line include phase-incoherent Cooper pairs (25-27), spin-singlet physics intertwined with CDW order (28-31) or pair-density wave fluctuations (32). Given that $\chi_{spin}$ of underdoped cuprates is better understood from a local moment picture (see discussion below), an interpretation of our results solely in terms of CDW-induced changes in the density of states (DOS) at $E_F$ is likely not tenable. The second class of explanation, or a combination of the two, may be more plausible options because the short-range CDW manifestly has an impact on AFM fluctuations. Indeed, it was recently realized that the decrease of low-energy AFM fluctuations upon cooling below ~120-150 K, long documented in Cu NMR relaxation and neutron scattering experiments in YBCO (4,34 and refs. therein) and initially interpreted as a signature of the pseudogap, is in fact related to the short-range CDW (14,15).

## The edge of the superconducting dome

We now contrast the results in the CDW regime with those at $p = 0.064$ ($YBa_2Cu_3O_{6.35}$, $T_c = 10$ K in zero-field) for which there is very little of CDW correlations, if any (13,14). Instead, weak incommensurate AFM glassy order is known to be present at low temperatures (35-37). This does not show up in Knight shift measurements because the hyperfine field produced by staggered moments cancels by symmetry at $^{89}$Y sites. However, spin freezing at $T_{spin} = 12$ K is detected in our dynamical measurements on this sample (Fig. S4).

As Fig. 2a shows, $\chi_{spin}$ is constant from base temperature to ~100 K, before rising upon warming towards room temperature. This dependence superficially looks different from that at higher doping but is again explained by the sum of a constant term (to be discussed later) and an exponential (see Fig. S1 for logarithmic plots *vs*. $T$ and $1/T$):

$K_{spin}(T) = K_{res} + K_H \exp(-\Delta_H/k_B T)$, (Eq. 3)

the difference being that the gap $\Delta_H \approx 600$ K is now considerably larger than $\Delta_L$ found at higher doping. Such a large gap $\Delta_H$ can neither be explained by the magnetic order, that merely appears as a low-temperature complication, nor by the $S$=1/2 Heisenberg model on the square lattice that shows decreasing, but not exponentially activated, $\chi_{spin}(T)$ (Fig. 1d and refs. 3,33). Since $\Delta_H \approx 600$ K essentially matches the pseudogap energy scale $\Delta_{PG}$ at this doping level (38), we conclude that the pseudogap is actually characterized by a well-defined energy gap for spin excitations at $q = 0$. Notice here that if we define $T^*$ as the temperature at which $\chi_{spin}$ reaches a maximum and starts to decrease upon cooling, $T^*$ should be around 800 K, larger than $\Delta_H/k_B$ (see ref. 33 and Supplementary Fig. S5 for data at $p \approx 0.14$). This illustrates that the initial decrease of $\chi_{spin}$ due to the onset of AFM correlations around $T^*$ and the gap are distinct effects. In fact, a pseudogap can lead to a decrease in $\chi_{spin}$ without any activated behavior at low temperature (Fig. 1d for an example in another context).

## Similarity with hole-doped spin ladders

The combination of an activated $\chi_{spin}(T)$ with a finite $\chi_{spin}(T=0)$ is uncommon in the absence of CDW order. It is found in electron-doped iron-based superconductors (41,42) but the activated behavior there is tied to the multiband electronic structure and is essentially doping-independent. This is therefore not relevant to cuprates. To the best of our knowledge, the only other similar case in a system that may be compared to YBCO is the quasi-1D cuprate $Sr_{14-x}Ca_xCu_{24}O_{41}$. With its nonmagnetic ground state and a large quantum spin-gap $\Delta$=0.4$J$ (Fig. 1b), the undoped ladder at x = 0 is an epitome of valence-bond physics, namely moments forming short-ranged singlets at temperatures where they would order if quantum fluctuations were insignificant. Hole-doping this material reduces the spin-gap, makes its infrared response quite similar to YBCO (43) but does not induce metallicity and $\chi_{spin}(T=0)=0$ up to x=12 (44). However, applying hydrostatic pressure at x=12 induces two-dimensional metallic transport (45) and, concurrently, a finite $\chi_{spin}(T=0)$. Still, the spin-gap remains visible in $\chi_{spin}$ (Fig. 1c and ref. 44).

Also similar to YBCO, is the weak AFM order actually present at low temperatures in the doped ladder (46,47). Despite the presence of these in-gap excitations, a spin-gap remains however discernible in the excitation spectrum, showing that singlet physics is still relevant in $Sr_{14-x}Ca_xCu_{24}O_{41}$ (x≈12). The coexistence of such AFM order with spin singlets is, in fact, a hallmark of gapped quantum antiferromagnets in the presence of quenched disorder, which is also relevant to YBCO (48-50).

In the doped ladder, the quantum spin-gap is clearly responsible for the activated behavior in $\chi_{spin}$. The remarkably similar $\chi_{spin}(T)$ in YBCO thus invites a comparable interpretation, signifying that the long-sought valence-bond physics is indeed relevant to the cuprates (68). Of course, since $\chi_{spin}$ measurements alone cannot definitively determine the origin of the gap, several objections may arise: an analogy does not constitute proof, a spin gap does not account for the pseudogap in the single-particle spectral function and the lack of prior evidence poses challenges. We thus now address these concerns in more detail.

The key point in our interpretation is that cuprates are doped Mott insulators and that $\chi_{spin}(T)$ should be approached from a local-moment perspective. While this has been established for $La_{2-x}Sr_xCuO_4$ (LSCO) (3,33), other cuprates like YBCO are often perceived as more itinerant. However, this is a misconception: at a given doping level, the $CuO_2$ planes of YBCO and LSCO have identical $\chi_{spin}(T)$, as Fig. S5 demonstrates for $p$=0.14 doping. At very low doping, $\chi_{spin}(T)$ is still very close to that of the undoped compound above its ordering temperature (33). It then appears extremely unlikely that the 600 K gap in $YBa_2Cu_3O_{6.35}$ could arise from anything other than a spin property (e.g. band-structure effect, preformed Cooper pairs or loop-current magnetism). In fact, for local moments, it is difficult to imagine a credible alternative to gapped singlets. Furthermore, theoretical works find that a spin gap can produce a gap of the same magnitude in the spectral function (see ref. 69 for a particularly explicit example), indicating no inconsistency between our interpretation and the pseudogap observed by photoemission in cuprates.

It should be made clear that we do not claim that underdoped YBCO is a gapped spin liquid as the analogy drawn above is *not* with the undoped ladder. The point is that there exists a metallic compound, composed of identical hole-doped elementary $CuO_2$ plaquettes, for which the coexistence of a quantum spin gap with gapless fermionic excitations produces the same unusual and surprisingly simple $\chi_{spin}(T)$ as found here in $YBa_2Cu_3O_{6.35}$ (Eq. 3). We thus argue that this analogy is both legitimate and compelling. Incidentally, the ladder material in question is also superconducting (45).

The lack of previous evidence of a gapped response may not be surprising. Its observation here relies on an exceptional combination of favorable factors converging in the strongly-underdoped regime of YBCO: $B_{c2}$ is low enough for steady magnetic fields to expose the normal state down to low temperatures, spin order is too weak to significantly impact on $\chi_{spin}$ (unlike striped cuprates) and the CDW onset temperature $T_{CDW}$ is much lower than the pseudogap temperature $T^*$ (unlike Bi2201), which is crucial to disentangle the different contributions to $\chi_{spin}$. Also, the relatively weak disorder in YBCO minimizes both the probability of broken singlets and the spatial inhomogeneity of magnetic properties. To the best of our knowledge, these conditions are not met simultaneously in any other cuprate.

In the doped ladder, the underlying gap appears to be more evident in the static response at $q$=0 (where it effectively adds to a constant background) than in the excitations at $q=Q_{AFM}$ where it is partially masked by in-gap states (46,47). Whether this also occurs in cuprates (60-62) remains an open question.

## Bridging the two gaps

We now discuss how to connect our two very different sets of observations: a high spin-gap $\Delta_H$ for $p$=0.064 and a much lower gap $\Delta_L$ for $p$=0.072 and above. It is extremely unlikely that $\Delta_H \approx 600$ K collapses with just a few percent of extra holes. The gap for $p$=0.072 was obtained by restricting the fit to temperatures below 80 K in order to capture the significant drop seen below 50 K (a single-gap fit over the entire $T$ range completely fails to capture the low $T$ drop and is thus inadequate). However, our fit (Fig. 2b) completely misses the high temperature data that actually shows a similar slope as the $p$=0.064 sample. This suggests a finite $\Delta_H$. We thus now fit the data with the sum of two gap functions:

$$K_{spin}(T) = K_{res} + K_H \exp(-\Delta_H/k_B T) + K_L \exp(-\Delta_L/k_B T). \qquad \text{(Eq. 4)}$$

Eq. 4 provides an excellent fit of the atypical, nearly-linear temperature dependence of $K_{spin}$ for $p$=0.072 (Fig. 2). This also holds true for $p$=0.09, although this doping is a less stringent test as the single-gap fit is already good.

Evidently, data from $p$=0.072 and $p$=0.09 alone would not conclusively establish evidence for two gaps. In fact, the compelling case for two separate gaps in this doping range stems from the unmistakable exponential behavior at $p$=0.064 combined with the likelihood of continuity as a function of doping. The overall consistency of results from single- and bi-exponential fits, further supports this conclusion. Specifically, the $\Delta_L$ values obtained from Eq. 4 are essentially the same as those derived from a single-gap fit to the low temperature data, while the obtained $\Delta_H$ values decrease about linearly with $p$ (Fig. 3a), which is consistent with the known trend of the pseudogap. Furthermore, the $K_{res}$ values agree with those deduced from independent measurements of the field dependence at low temperatures (see Fig. S6).

Upon raising $p$ beyond 0.10, however, the two-gap fit becomes unstable, forcing us to fix $\Delta_H$ values by linear extrapolation of the data at $p \leq 0.09$ (crosses in Fig. 3a). These biexponential fits are visually indistinguishable from single-exponential fits (and are thus not shown). They yield identical $\Delta_L$ values as obtained from single-gap fits and an exponential weight $K_H$ that continues to decrease with $p$, until eventually vanishing, whereas $K_L$ continuously increases (Fig. 3b and Fig. S3). The data is thus consistent with $\Delta_H$ being still present (*i.e.* there are two gaps) up to at least p=0.135. The continuous decrease of the weight $K_H$ with $p$ may either reflect a true physical effect (see discussion in the last section) or an artifact of the fit converging towards the lowest gap value as $\Delta_H$ becomes too close to $\Delta_L$. In any event, the results confirm our initial conclusion that the lowest gap is related to the CDW, not to the pseudogap.

## Residual spin susceptibility

As mentioned above, $K_{spin}$ does not vanish at $T = 0$: there is a residual shift $K_{res} \propto \chi_{spin}(T{=}0)$ that decreases upon increasing $p$ from 0.064 to 0.09 and then shows little or no variation up to $p$=0.135 (Fig. 3c).

A finite $\chi_{spin}(T{=}0)$ may arise from a finite density of triplets in an otherwise singlet ground state (39) or from gapless magnetic excitations. On the other hand, that $K_{res}$ enters Eq. 4 as a temperature-independent term is also reminiscent of the Pauli susceptibility of a simple metal, which is proportional to the DOS. The magnitude of $\chi_{spin}(T{=}0)$ does not deviate unreasonably from the value expected for free electrons: using data from ref. 67, we convert $K_{res}$=0.024% into $\chi_{spin}(T{=}0)$ = 3.7 $10^{-5}$ emu $mol^{-1}$ and with an electronic specific heat $C_{el}/T$ of 4.8 mJ $mol^{-1}$ $K^{-2}$ in the $T$=0 limit in high fields from ref. 40 (a number that decreases from $p$=0.10 to $p$=0.12, thus aligning with $K_{res}$), we calculate a Wilson ratio $R_W \approx 0.6 \pm 0.2$ for $p$=0.11. However, that $\chi_{spin}(T{=}0)$ decreases with doping does not fit with expectations for a contribution of metallic type. This implies that $\chi_{spin}(T{=}0)$ is either unrelated to simple particle-hole excitations or it represents only a part of a larger metallic contribution that is $T$ dependent (for instance due to partial gapping by the CDW, see Figs. 1e and 3d). While the current data already provide quantitative constraints, more work is surely needed to work out the nature of the excitations responsible for the residual susceptibility.

These results sharply contrast with those obtained from Bi2201 in the only comparable high-field NMR study conducted thus far. There, $K_{res}$ values are much larger and they continuously increase with doping (7). The most obvious difference between YBCO and Bi2201 is the level of disorder. However, we find that our $K_{res}$ values do not correlate with the level of disorder deduced from the NMR linewidth (Fig. S7), which indicates that $\chi_{\mathrm{spin}}(T=0)$ is mostly of intrinsic origin here. Regardless of microscopic interpretations, it is conceivable that the more developed CDW correlations in YBCO (~3 times longer CDW correlation length at $p\approx0.12$) act to reduce $\chi_{\mathrm{spin}}$ with respect to Bi2201.

## Discussion and outlook

By measuring the spin susceptibility of underdoped $YBa_2Cu_3O_y$ in high fields, we have discovered that it is gapped at low temperatures, even though superconductivity has been suppressed. We found two distinct activated contributions and have attributed them to spin-singlet and CDW correlations. The following expression captures $\chi_{\mathrm{spin}}$ at low temperatures:

$$\chi_{\mathrm{spin}}(T) = \chi_{\mathrm{res}} + \chi_1 \exp(-\Delta_{\mathrm{CDW}}/k_B T) + \chi_2 \exp(-\Delta_{\mathrm{PG}}/k_B T)\ , \qquad \text{(Eq. 5)}$$

where $\Delta_{\mathrm{CDW}} \approx k_B T_{\mathrm{CDW}}$ is an energy scale associated with CDW formation (but not necessarily the CDW gap) and $\Delta_{\mathrm{PG}}$ is the known pseudogap energy scale. This equation, as simplified as it may be (notably ignoring any possible $T$ dependence of $\Delta_{\mathrm{CDW}}$), will serve as a benchmark for future calculations of $\chi_{\mathrm{spin}}$ at low $T$, especially as finite-temperature algorithms now start to reach the static CDW regime (51). Furthermore, it offers valuable insight into the pseudogap state of YBCO and raises a number of important questions, as we now discuss.

An interesting issue is whether Eq. 5 can be accounted for by a single electronic fluid (52). The presence of two separate gaps appears to require some form of electronic differentiation (30,33, 53-55) but this could as well occur in reciprocal space (nodal and anti-nodal states of the one-band Hubbard model, for instance) as in real space (O2$p$ and Cu3$d$ bands). High-field $^{63}$Cu NMR experiments would be desirable to address this issue but these are exceedingly challenging in YBCO due to the accidental vanishing of $^{63}K_{\mathrm{spin}}$ in perpendicular fields (49,52).

We have interpreted Eq. 5 as evidence that the pseudogap in $\chi_{\mathrm{spin}}$ is contributed by spin singlets and, when present, by short-range CDW order. This adds to the initial effect of short-range AFM correlations at $T \approx T^*$ (that are not described by Eq. 5, as discussed previously). Thus, different phenomena contribute to the pseudogap in a single physical property, with a relative influence that depends upon temperature and doping. This suggests that the pseudogap should be viewed as a composite property, a notion that conciliates seemingly differing viewpoints on the pseudogap problem. In particular, that CDW correlations reduce $\chi_{\mathrm{spin}}$ at $T << T^*$ should not be regarded as contradictory to the notion that $\chi_{\mathrm{spin}}$ starts to decrease at $T \approx T^*$ because of the onset of AFM correlations.

From a theoretical perspective, the field has come a long way since Anderson's seminal proposal of gapless singlets at half filling (56). A number of ideas potentially explain the emergence of a spin-gap upon doping (39,57 for example) but in many of them this occurs in tandem with CDW order (28-30,32). This is certainly relevant to our results in the CDW regime. Nonetheless, our data at doping $p \leq 0.08$ reveal another spin gap, not related to CDW order (in addition to gapless excitations). This points to the relevance of valence-bond physics for understanding (part of) the spin excitations in the normal state. How this physics fades away upon increasing hole doping is however unsettled. The spin-gap presumably decreases until it vanishes at a certain doping level but the notion of singlets and triplets may also become increasingly ill-defined. Taken at face value, the opposite but continuous dependence of the exponentials' weights from $p$=0.072 up to 0.135 (Fig. 3b) could be evidence that the local moments get dissolved into the metal in a progressive fashion. However, as cautioned above, one should not overinterpret the fit results and, to our view, this issue remains open. In this regard, a relevant question is whether singlets could be particularly favored either in YBCO because of its orthorhombicity and/or more generically at low doping due to the kind of plaquette physics recently discovered in Bi2201 (58,59).

Finally, it is interesting to place the results in the context of the recent progress in numerical studies of the single-band Hubbard model on the square lattice: this minimal proxy for a $CuO_2$ plane is now widely recognized to have a CDW ground state around $p \approx 0.12$ (63,64). However, there remains a major discrepancy between theory and experiments: numerical studies almost systematically find spin-stripe order intertwined with the CDW. Experiments on the other hand, find spin-stripes only in La-based cuprates and spin order is notably absent in YBCO even when the CDW is long ranged and unidirectional (10). Still, we have argued that the same local moments as in LSCO are present in YBCO. The question then arises as to whether the enhanced propensity to form singlets in YBCO could be related to its CDW periodicity of three lattice units (13,14,65) instead of the typical four in LSCO. In this regard, we notice the recent discovery (66) of a non-magnetic, period-3 CDW (so-called W3 phase) at low doping in the $t$-$t'$-$J$ model ($t'/t \approx$-0.2), which strikingly matches our observations in YBCO. This begs the question as to whether distinctive properties of YBCO (such as orthorhombicity, bilayer coupling and/or a different value of $t'/t$) help to stabilize the W3 phase at higher doping levels.

Answering the above questions can aid in identifying the decisive factors that tilt the balance towards either spin-stripe order or spin singlets. Ultimately, this endeavor has the potential to illuminate the elusive origin of cuprate superconductivity.

[a]Present address: Institute of Physics, Chinese Academy of Sciences, and Beijing National Laboratory for Condensed Matter Physics, Beijing 100190, China.

[b]Present address: IV Physics Institute, University of Göttingen, 37077 Göttingen, Germany.

[c]Present address: MPA-Q, Los Alamos National Laboratory, Los Alamos, 87545 New Mexico, USA.

[d]Present address: Hefei National Research Center for Physical Sciences at the Microscale, University of Science and Technology of China, Hefei, China

**Acknowledgments**

We thank K. Behnia, C. Bernhard, P. Bourges, A.V. Chubukov, A. Georges, S.A. Kivelson, D. LeBoeuf, P. Lee, P. Mendels, D. Orgad, C. Proust, M. Punk, S. Sachdev, A. Sacuto, J. Schmalian, Y. Sidis, J. Tallon, J.M. Tranquada, A.-M. Tremblay, S.R. White, J. Zaanen for discussions.

This work was performed at the LNCMI, a member of the European Magnetic Field Laboratory. Work in Grenoble was supported by the Laboratoire d'Excellence LANEF (ANR-10-LABX-51-01) and by the French Agence Nationale de la Recherche (ANR) under reference ANR-19-CE30-0019 (Neptun).

**Data availability**

The data that support the plots within this paper are available as source data files. All other data are available from the corresponding author upon reasonable request.

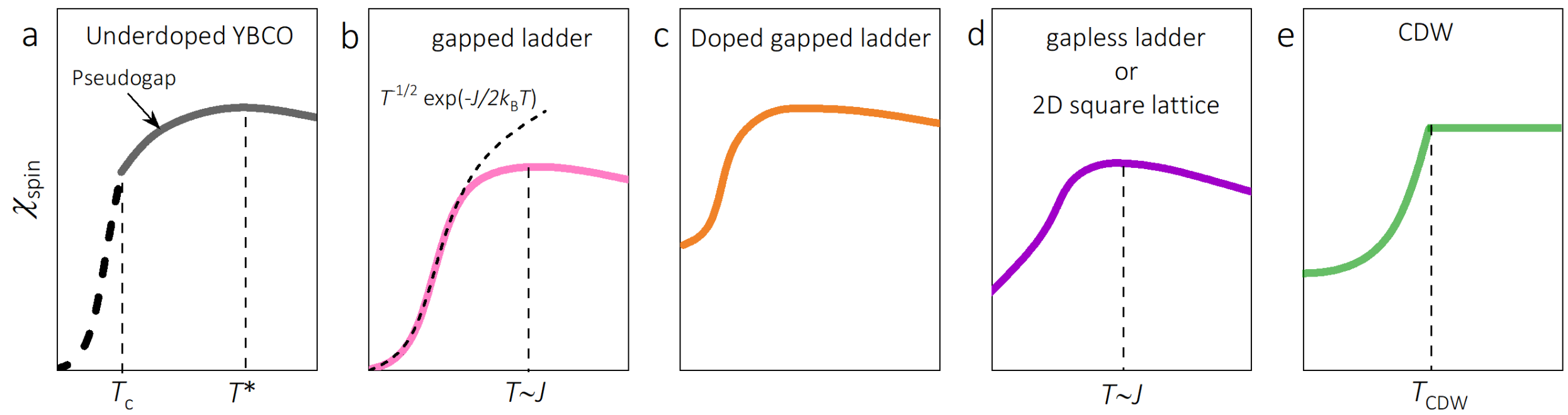

**Fig. 1. Examples of systems with gaps or pseudogaps in the spin susceptibility.** (a) Typical pseudogap behavior in underdoped YBCO. $\chi_{spin}$ drops below $T_c$ because of singlet superconducting pairing. The pseudogap corresponds to the decrease upon cooling below $T^*$. (b) Gapped behavior of an even-legged spin ladder. For two legs and equal Heisenberg exchange $J$ between rungs and legs, the spin-gap $\Delta = J/2$. The dashed curve represents the low $T$ approximation $\chi_{spin} = \chi_{res} + \chi_0\ T^{-1/2} \exp(-\Delta/k_B T)$ (6). AFM correlations develop below $T \sim J$, making $\chi_{spin}$ decrease upon cooling already well above the scale $T \sim J/2$ of singlet formation. (c) $\chi_{spin}$ of the hole-doped two-leg ladder $Sr_2Ca_{12}Cu_{24}O_{41}$ in its 2D conducting phase under pressure, with an exponential decrease above a finite-$\chi$ background (44). (d) Typical behavior of an odd-legged spin ladder or a 2D Heisenberg square lattice, both lacking a spin gap (3,6,33). AFM correlations make $\chi_{spin}$ decrease below $T \sim J$. (e) Typical $\chi_{spin}$ of some CDW systems (23,24).

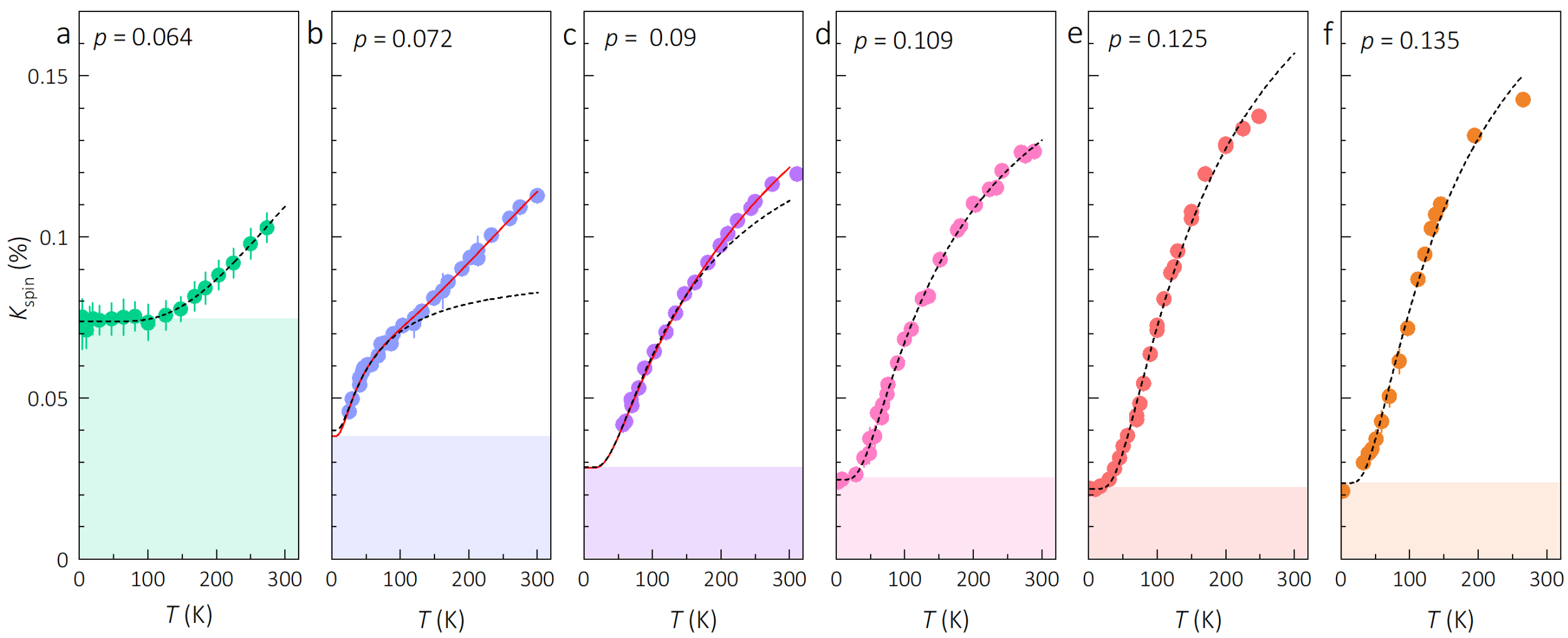

**Fig. 2. Temperature and doping dependence of the spin susceptibility.** $K_{spin}$ is the spin part of the Knight shift, which is proportional to $\chi_{spin} = \chi'_{spin}(q=0,\omega=0)$ (see Eq. 1). As described in SM, the datasets result from a variety of measurements in different orientations of the field, using $^{17}O$ NMR ($^{89}Y$ NMR for $p$=0.064) and they have been rescaled in equivalent $^{17}K_{spin}$ values for O(2) sites with $H||b$ axis. Dashed curves represent single-exponential fits (Eqs. 2 & 3) and red solid curves denote bi-exponential fits (Eq. 4). For panels d,e,f, bi-exponential fits are virtually indistinguishable from single-exponential fits. The colored background represents the residual contribution $K_{res} = K_{spin}(T=0)$ (Eqs. 2, 3). As shown in SM, the $K_{res}$ value for $p$=0.072 corresponds to the value expected on the basis of the field dependence at low $T$.

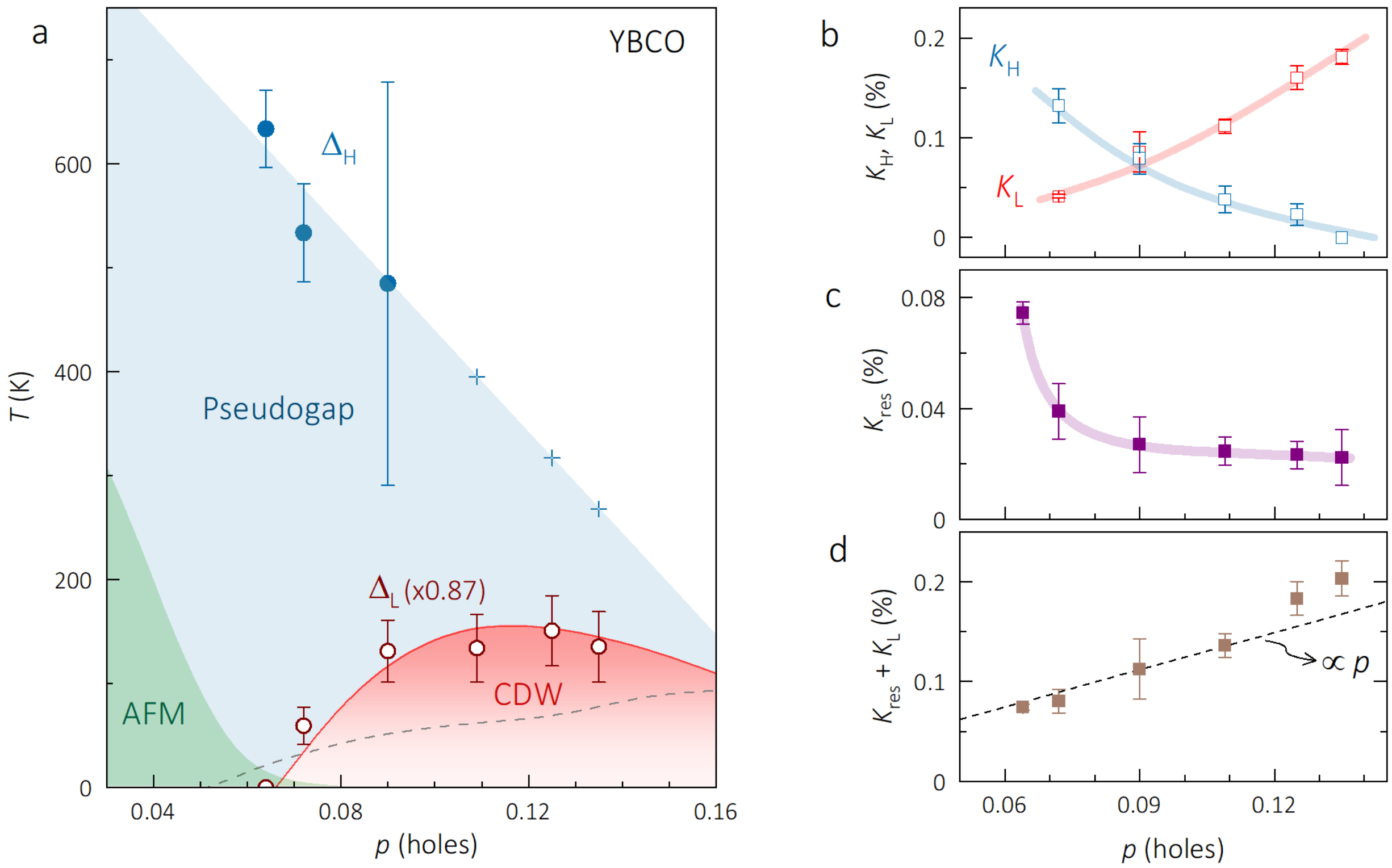

**Fig. 3. Two-gap phenomenology and phase diagram of underdoped $YBa_2Cu_3O_y$.** (a) Temperature – doping phase diagram with the two gap values $\Delta_H$ and $\Delta_L$ extracted from fits (see text). Crosses represent $\Delta_H$ values fixed in the fit (from linear extrapolation of data at lower doping). The red line is a guide to the eye through data for $T_{CDW}$ (the onset temperature of the short-range CDW), as determined by x-ray and NMR (see SM). The dashes represent the zero-field $T_c$. (b) Pre-exponential factors $K_H$ and $K_L$ (see Eq. 4). (c) Residual Knight shift $K_{res} = K_{spin}(T=0)$ due to gapless excitations. (d) The sum of $K_{res}$ and $K_L$ increased (initially linearly) with hole doping. All data correspond to the average values of fit results with exponential and hyperbolic tangent functions (both give very similar results, see SM). Error bars for $\Delta_H$, $K_H$, $K_L$ and $K_{res}$ represent one standard deviation in the fit results. Error bars for $\Delta_L$ represent the uncertainty related to different possible fit functions (the error bar for each fit is about the symbol size, see SM).

# Methods

## Samples

High-quality, oxygen-ordered, detwinned single crystals of $YBa_2Cu_3O_y$ were grown in non-reactive $BaZrO_3$ crucibles from high-purity starting materials. For ortho-II ordered samples, the oxygen content was determined from the low-frequency quadrupole satellites of the $^{63}$Cu NMR spectrum in oxygen-empty chains, according to the method described in ref. 69 (see Fig. S8). Supplementary Table 1 lists the studied crystals and their properties.

## Experimental methods

Magnetic fields were provided by superconducting magnets up to 15 T and by the M9 and M10 resistive magnets of LNCMI Grenoble for higher field values. Standard spin-echo techniques were used with a laboratory-built heterodyne spectrometer. Spectra were obtained at fixed magnetic field by adding Fourier transforms of the spin-echo signal recorded for regularly spaced frequency values. For $^{17}$O-NMR measurements, the field values were calibrated using the resonance frequency of the central line of $^{17}$O(4E), *i.e.* apical oxygen sites located below oxygen-empty chains, for which $K = 0$ within error bars. For $^{89}$Y-NMR measurements, the coil was made with silver wire, and $^{109}$Ag-NMR was used to calibrate the magnetic field with $^{109}K = 0.525\%$ at room temperature.

## Data analysis

Three different types of datasets were recorded:

1) At temperatures above the zero-field $T_c$, the $^{17}$O Knight shift $K$ was measured at O(2) sites (those planar sites in bonds oriented along the $a$ axis), with the field applied along the $b$ axis. This configuration maximizes the resolution (single O(2) site and minimal overlap with lines from other O sites in the NMR spectrum).

2) At low temperatures and high fields, $K$ was measured with the field tilted by 16° off the $c$ axis. The strong $c$ axis component is required to quench superconductivity but, without tilt angle, O(2) and O(3) sites would overlap and thus degrade precision. We measured the shift of O(3) sites (planar O sites in bonds oriented along the $b$ axis) because the spectral changes in the long-range CDW state are less complex for O(3) than for O(2) (this latter is affected by equally strong effects in both the quadrupole and magnetic hyperfine channels, which leads to very asymmetric NMR spectra). This is especially important as accurate determination of $K$ requires precise measurements of the quadrupole satellites.

3) For the $p$=0.064 sample, the shift of $^{89}$Y was measured because the $^{17}$O signal is wiped out at low $T$ by glassy magnetic order (seen as a broad peak in $1/T_1$ of $^{89}$Y shown in Fig. S4).

In order to determine the relationship between $^{89}K_c$ and $^{17}K_b$, we also measured both $^{89}K_c$ and $^{17}K_b$ for $p$=0.072, as shown in Fig. S9.

Pairs of these datasets were then scaled against one another (Figs. S9 and S10) in order to convert all data into equivalent $^{17}K_\mathrm{b}$ values ($^{17}$O(2) Knight shift with $B||b$). The measured total Knight shift is the sum of the spin and orbital contributions:

$K = K_\mathrm{spin} + K_\mathrm{orb}$

In the cuprates, $K_\mathrm{orb}$ is due to Van-Vleck paramagnetism and is $T$ independent. As Fig. S11 shows, the $K_\mathrm{orb}$ values are found to be very small at all doping levels. Furthermore, these values depend on $p$ non-monotonously (even changing sign). We thus conclude that $^{17}K_\mathrm{orb}$ = 0 within experimental uncertainty. Therefore, $K_\mathrm{spin}$, the spin part of the Knight shift (Fig. 2) is equal to the measured total Knight shift.

**Fitting procedure**

We fitted the $T$ dependence of the Knight shift with the following two expressions (or one-gap version thereof):

$K_\mathrm{spin} = K_\mathrm{res} + K_\mathrm{L} \exp(-\Delta_\mathrm{L}/k_\mathrm{B}T) + K_\mathrm{H} \exp(-\Delta_\mathrm{H}/k_\mathrm{B}T)$ (“exp” symbols in Fig. S3)

$K_\mathrm{spin} = K_\mathrm{res} + K_\mathrm{L} (1-\tanh^2(\Delta_\mathrm{L}/2k_\mathrm{B}T)) + K_\mathrm{H} (1-\tanh^2(\Delta_\mathrm{H}/2k_\mathrm{B}T))$ (“tanh” symbols in Fig. S3)

The second formula has been used previously in the cuprate context (70), based on a phenomenological model accounting for neutron scattering data in YBCO (71). It indeed fits slightly better (see Fig. S2) than the simple exponential function that is a bit steeper. Notice that we found the function $K_\mathrm{spin} = K_\mathrm{res} + K_0\, T^{-1/2} \exp(-\Delta/k_\mathrm{B}T)$, appropriate for even-leg spin ladders, to yield results that are essentially indistinguishable from those without the $T^{-1/2}$ pre-exponential factor.

| **Nominal O content** | **Hole doping (*p*)** | **Oxygen order** | **Measured nucleus** | $T_c$ **(*B*=0)** | $B_{c2}$ **(*T*→0 K)** | $B_{max}$ | **Ref.** |
|---|---|---|---|---|---|---|---|
| 6.38 | 0.064 | O-II | $^{89}$Y | 10 K | <20 T | 15 T | |
| 6.43 | 0.072 | O-II | $^{89}$Y, $^{17}$O | 37.8 K | ~45 T | 30 T | |
| 6.49 | 0.09 | O-II | $^{17}$O | 52.3 K | ~40 T | 15T | (9) |
| 6.56 | 0.109 | O-II | $^{17}$O | 59.8 K | ~24 T | 28.5 T | (12,9,61) |
| 6.68 | 0.125 | O-VIII | $^{17}$O | 67.8 K | ~24 T | 28.5 T | (12,9) |
| 6.77 | 0.135 | O-III | $^{17}$O | 78.3 | ~34 T | 34 T | |

**Supplementary Table 1**: **Sample properties.** The y=6.49 sample was initially labeled as y=6.47 in ref. (9) but its oxygen concentration was later reevaluated after measuring Cu(1E) NMR spectra. $B_{c2}(T=0)$ values are obtained by interpolating the results in refs. (72,73).

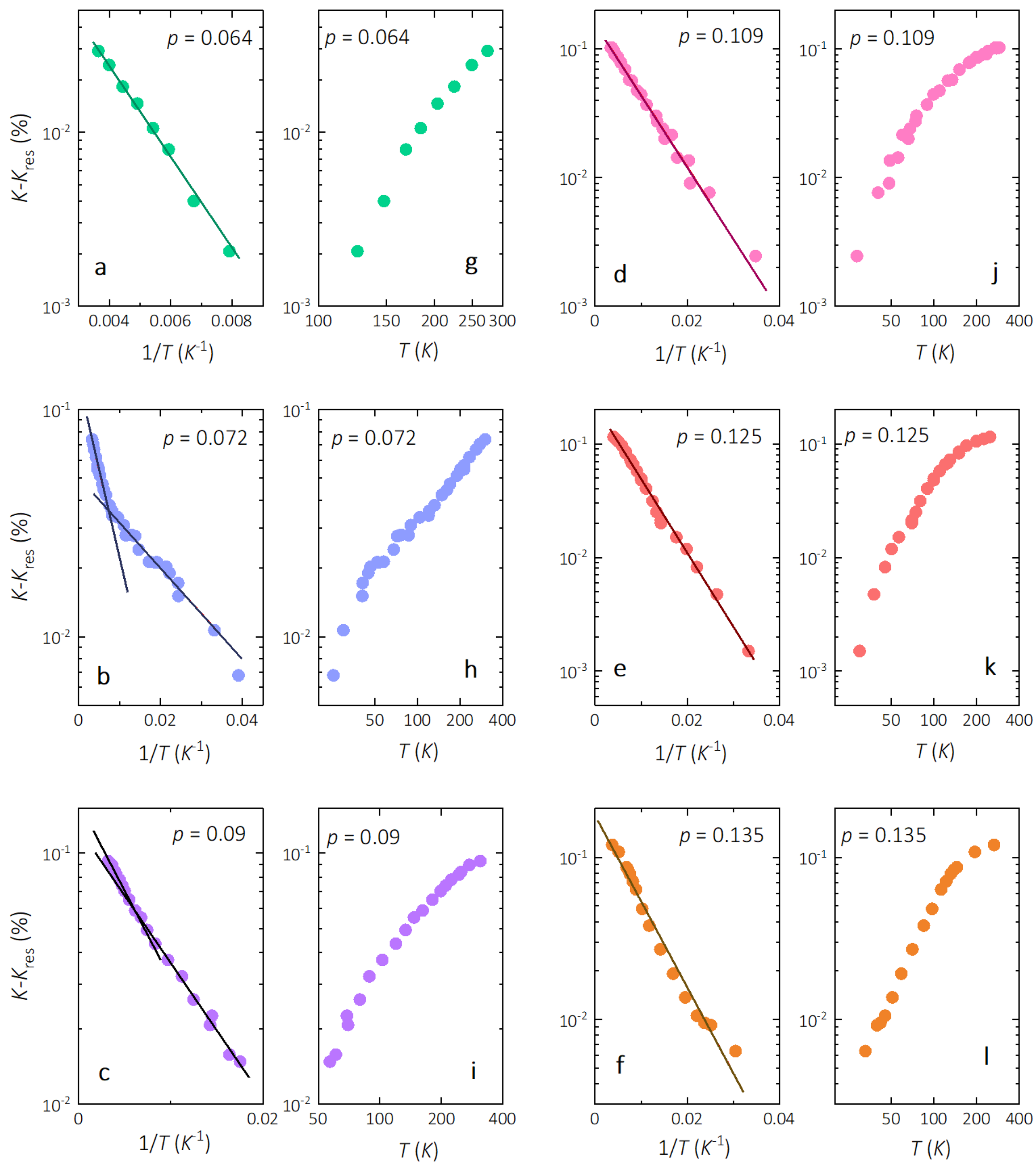

**Fig. S1. Evidence of activated behavior. a-f**: Knight shift (from which $K_{res}$ has been subtracted) in vertical logarithmic scale *vs.* inverse temperature for the different doping levels. The straight line indicates exponential behavior. **g-l**: same data *vs.* temperature. The nonlinear behavior speaks against a power-law dependence. The *p*=0.135 data undulates around the straight line (panel **f**) because of the larger difficulty in determining the exact line position for this sample: the line shapes are more complex due to ortho-III chain-oxygen order being both more complex and shorter ranged.

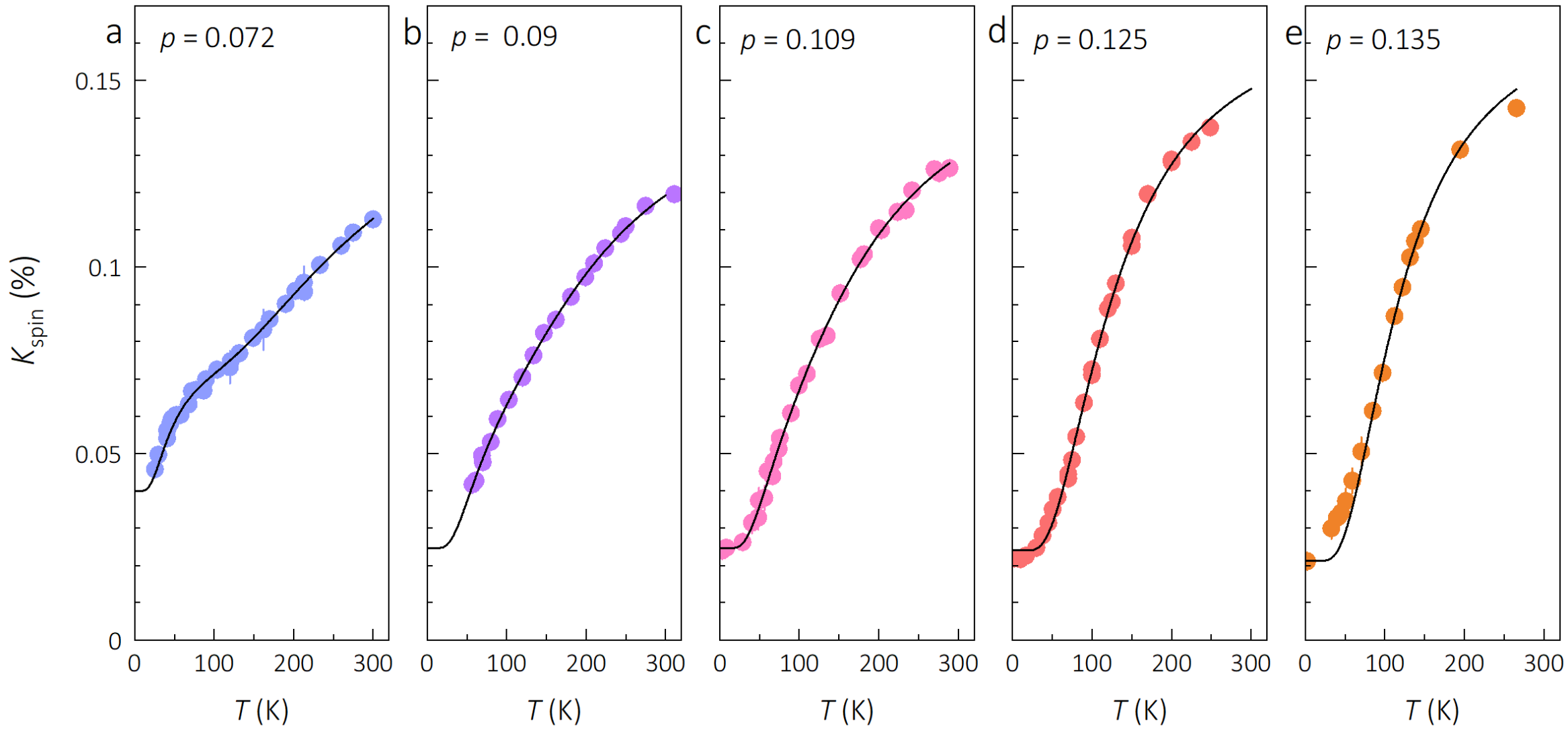

**Fig. S2. Alternative fits**. Same data as in Fig. 2. The solid lines are fits with the two-gap function $K_{\mathrm{spin}} = K_{\mathrm{res}} + K_{\mathrm{L}}\ (1\text{-}\tanh^2(\Delta_{\mathrm{L}}/2k_{\mathrm{B}}T)) + K_{\mathrm{H}}\ (1\text{-}\tanh^2(\Delta_{\mathrm{H}}/2k_{\mathrm{B}}T))$. Fit results are shown in Fig. S3.

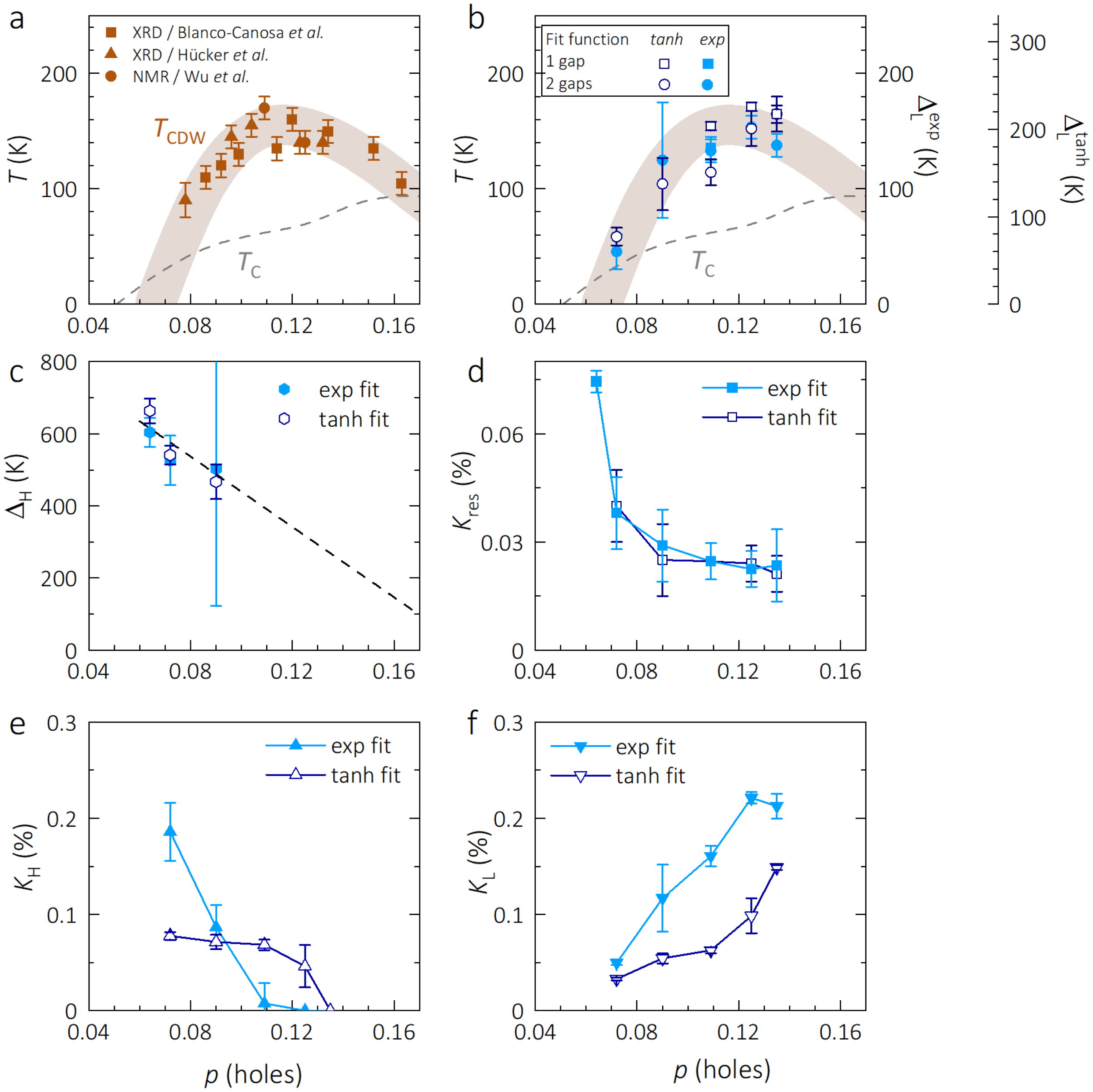

**Fig. S3. Comparison of results for two different fitting functions.** (a) Onset temperature $T_{CDW}$ of short-range 2D CDW, from refs. (13,14,15). The thick trace is a parabolic function that represents the $p$ dependence of $T_{CDW}$ and its experimental uncertainty. (b) Low-gap values $\Delta_L$ extracted from the two different fits, and using one or two gap functions (see text), compared to the same parabola as in (a). Error bars for $\Delta_H$, $\Delta_L$, $K_H$, $K_L$ and $K_{res}$ represent one standard deviation in the fit results.

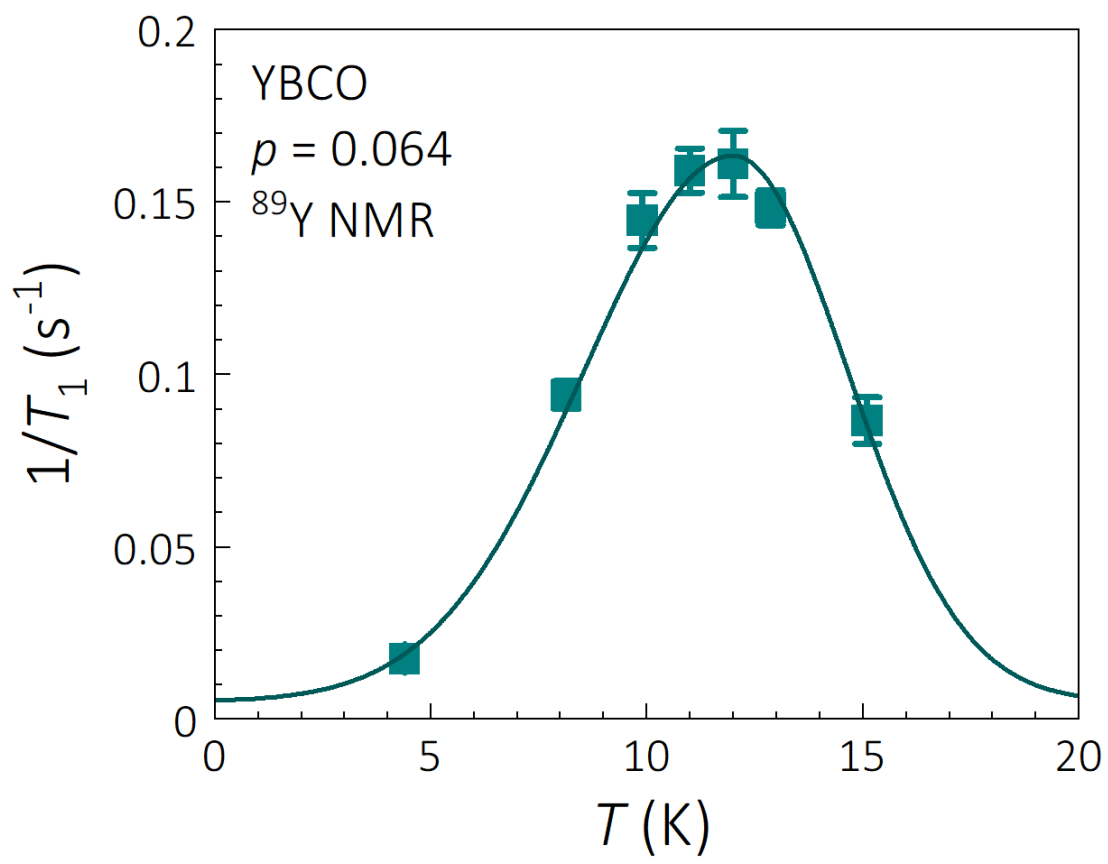

**Fig. S4. $^{89}$Y NMR evidence of spin freezing for *p*=0.064 ($YBa_2Cu_3O_{6.38}$).** A broad peak in the spin-lattice relaxation rate $1/T_1$ *vs.* *T* is the typical signature of spin fluctuations becoming as slow as the NMR frequency scale of 31 MHz (here in a field of 15 T).

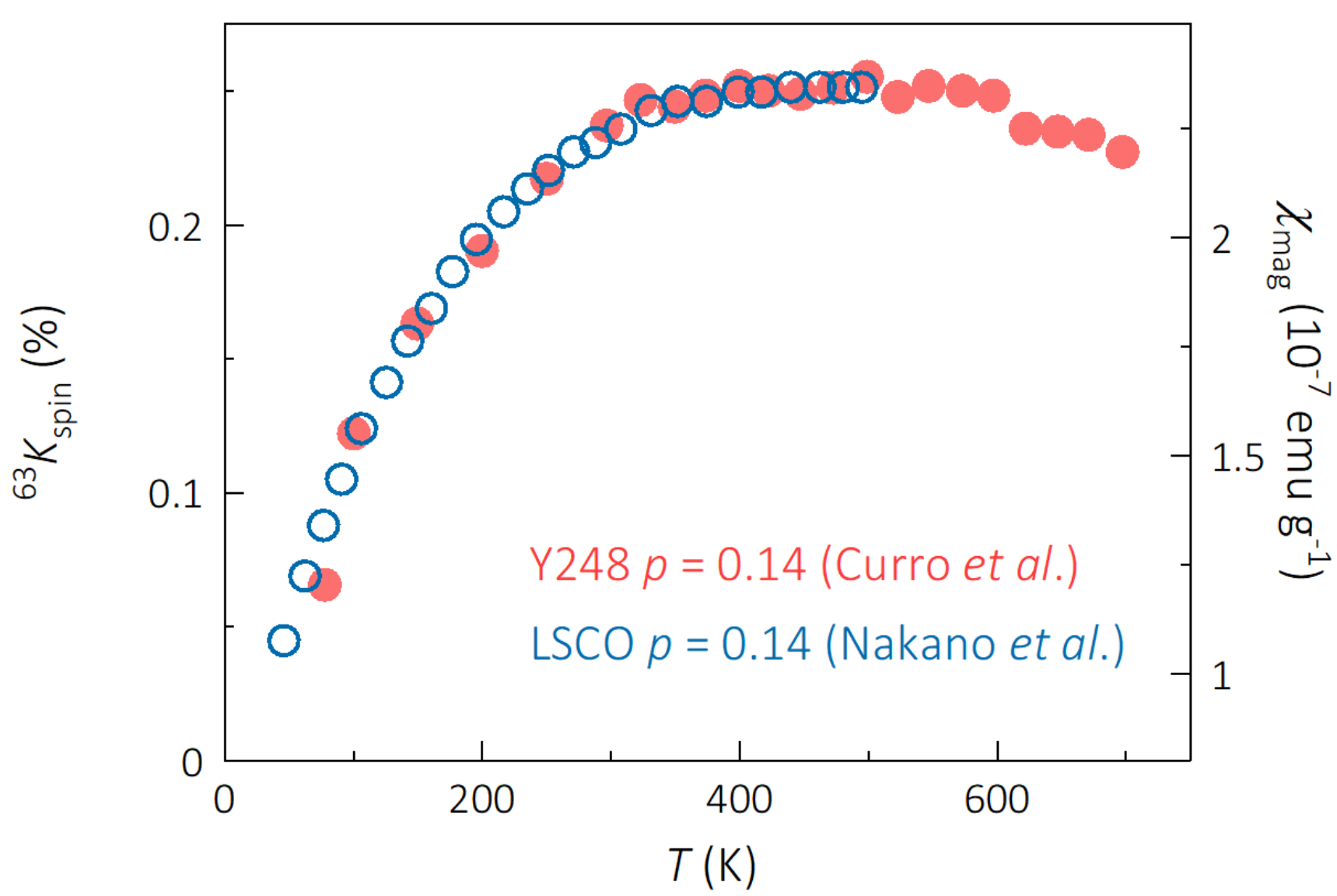

**Fig. S5. Comparison of the spin susceptibility of LSCO and YBCO.** Bulk magnetization data for $La_{1.86}Sr_{0.14}CuO_4$ is from Nakano *et al.* (33). $^{63}$Cu Knight shift data for $YBa_2Cu_4O_8$ is from Curro *et al.* (74). Unlike $YBa_2Cu_3O_y$ for which chain-oxygen atoms become mobile above room temperature, stoichiometric $YBa_2Cu_4O_8$ does not suffer this problem. This allows one to see that the broad susceptibility maximum, inherited from the half-filled 2D square lattice, is present in the YBCO system as well.

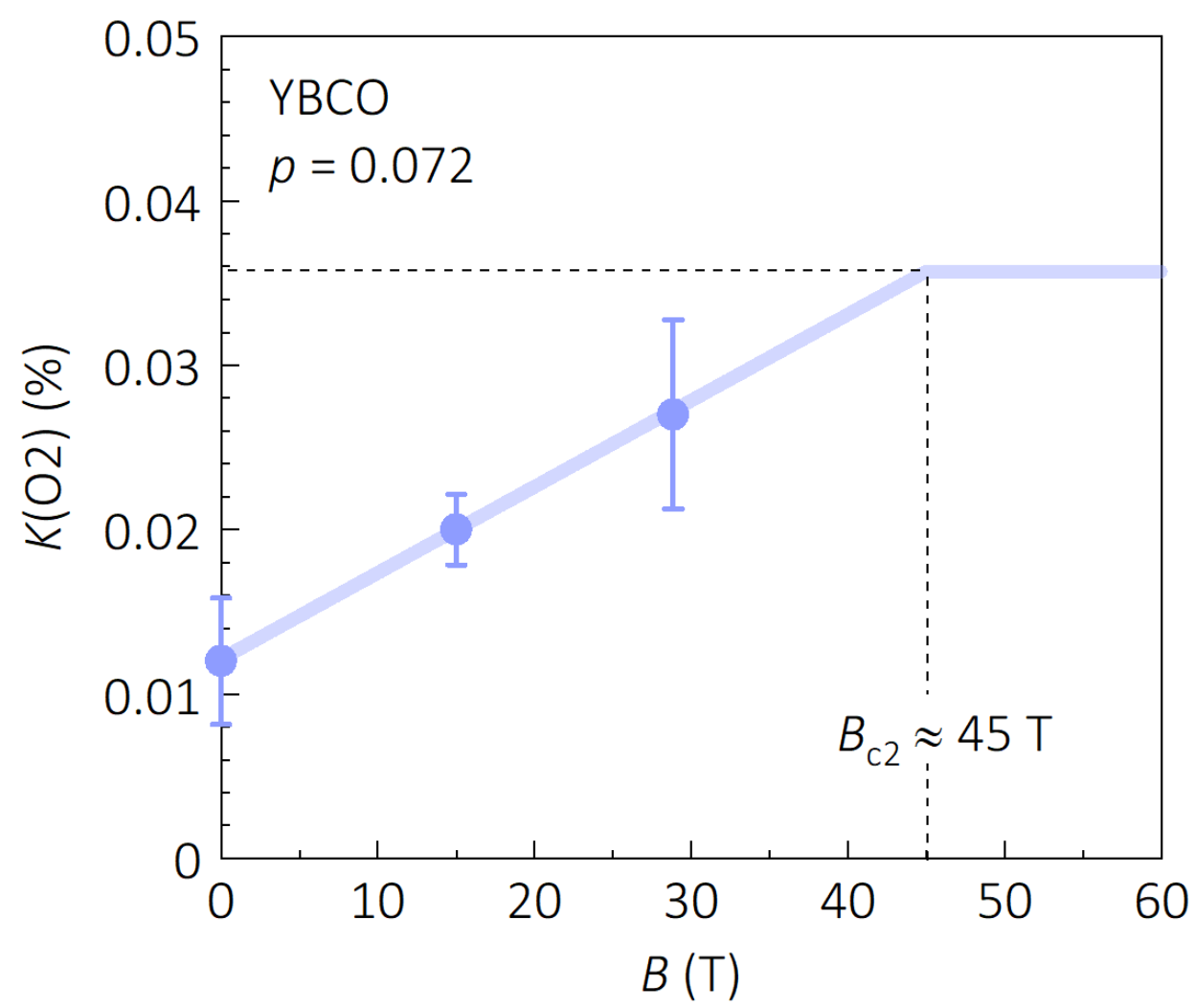

**Fig. S6. Predicting $K_{res}$ for $YBa_2Cu_3O_{6.43}$.** The data at $B$ = 15 T (measured with $B||c$) and at 30 T (measured with $B$ tilted by 16° off the $c$ axis) correspond to the zero-temperature extrapolation of our $K(T)$ measurements of O(2) sites at low $T$. The value at $B$ = 0 is the zero-temperature extrapolation of our $K(T)$ measurements of O(2) sites at low $T$, with $B \perp c$ and $B$ = 9T << $B_{c2}^{\perp}$. The thick blue line represents the expected $B$ dependence on the basis of results at other doping levels (9): linear increase up to $B_{c2}(T=0)$ = 45 T (value taken from an interpolation of the results in refs. 72,73) and saturation above $B_{c2}$. The value $K(T=0, B=B_{c2}) = K_{res}$ is predicted to be equal to about 0.036%. Within error bars, this agrees well with the fit result in Fig. 2b: $K_{res}$(fit) = 0.039±0.01%.

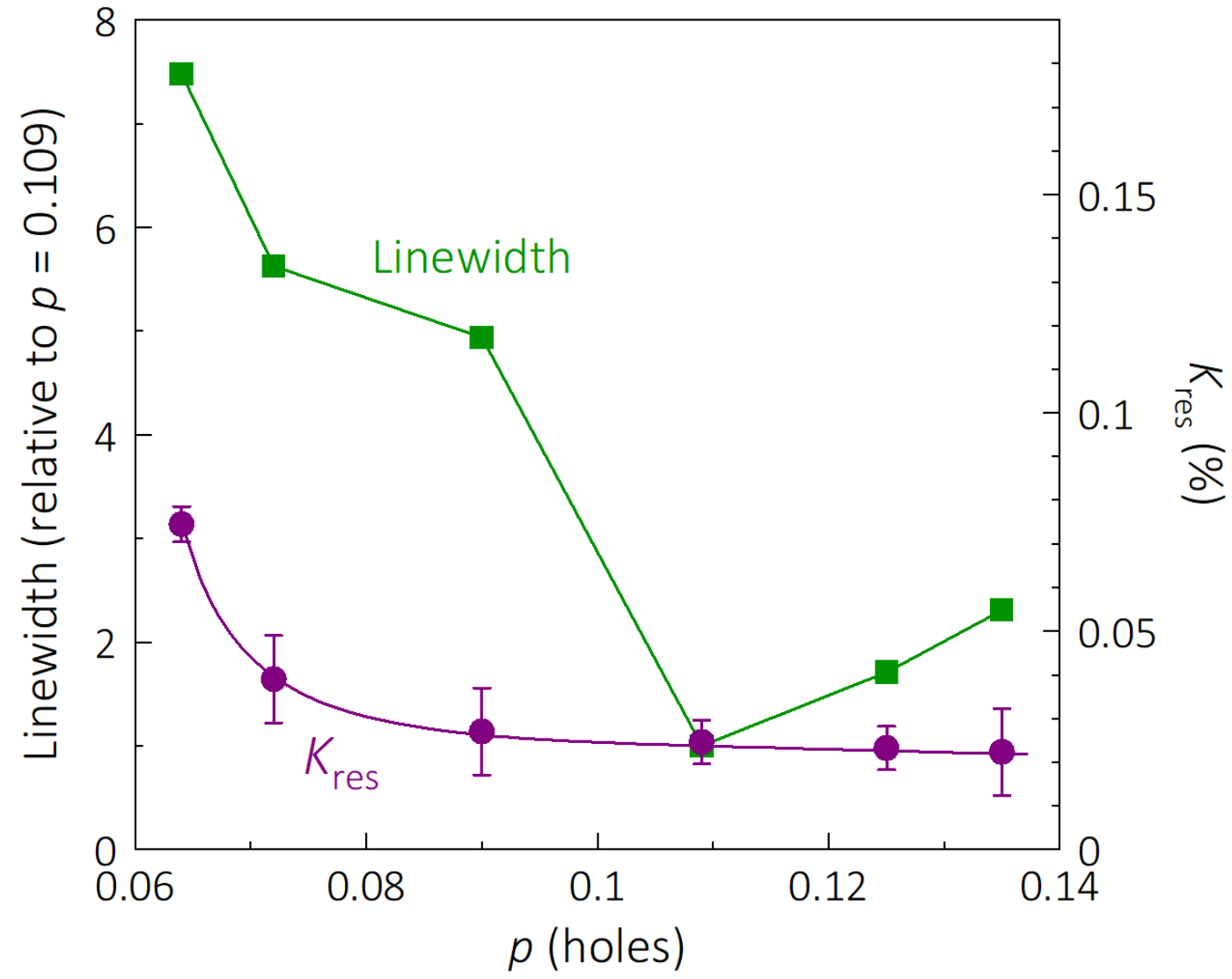

**Fig. S7. Absence of direct correlation between residual spin susceptibility and disorder.** The linewidth values correspond to the width of the $^{17}O(2)$ central line at room temperature. For $p$=0.064, we used the ratio of $^{89}Y$ NMR linewidth between $p$=0.064 and $p$=0.072 samples. While $K_{res}$=$K_{spin}(T=0)$ is nearly identical for $p$=0.109 and $p$=0.09, disorder (as quantified by the NMR line width) differs by a factor of ~5. $p$=0.135 also has identical $K_{res}$, yet its linewidth is larger than p=0.109 by a factor 2.3. Therefore, although both $\chi_{res}$ and disorder tend to increase at low doping, this likely occurs for different reasons: weakening of CDW correlations for the former, increased oxygen disorder for the latter (notice that the weakening of CDW correlations may, or may not, be partially caused by the increased disorder at low $p$).

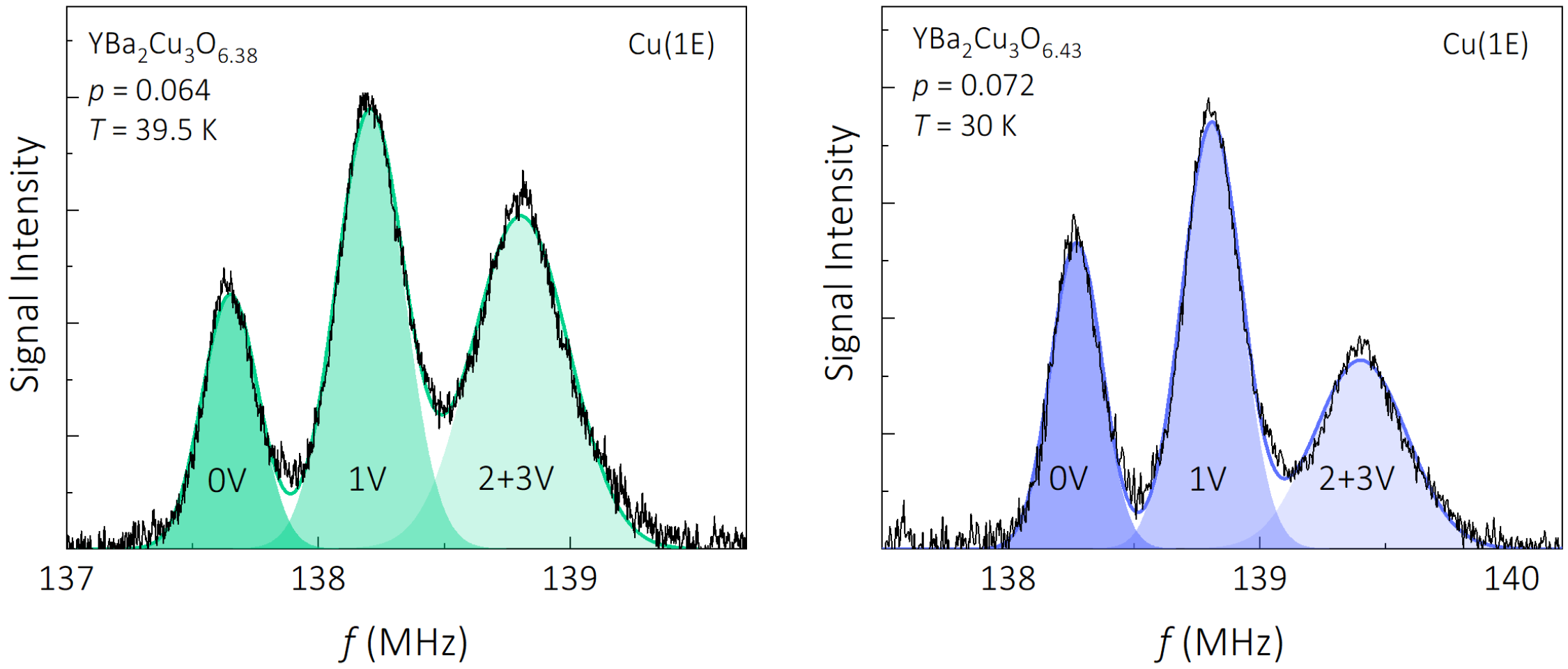

**Fig. S8: determining the oxygen concentration with chain-Cu NMR.** $^{63}$Cu(1E) (empty-chain site) low-frequency quadrupole satellites of the $^{63}$Cu NMR spectrum. The labels iV (i = 0, 1, 2, 3) indicate Cu(1E) sites having a number i of nearest-neighbor vacancies. Each line is fitted with a Gaussian function and the relative integrated intensities of the Cu(1E)$_{0V}$ and Cu(1E)$_{1V}$ sites are used to determine the actual oxygen concentration y (see ref. 69 for details about the method).

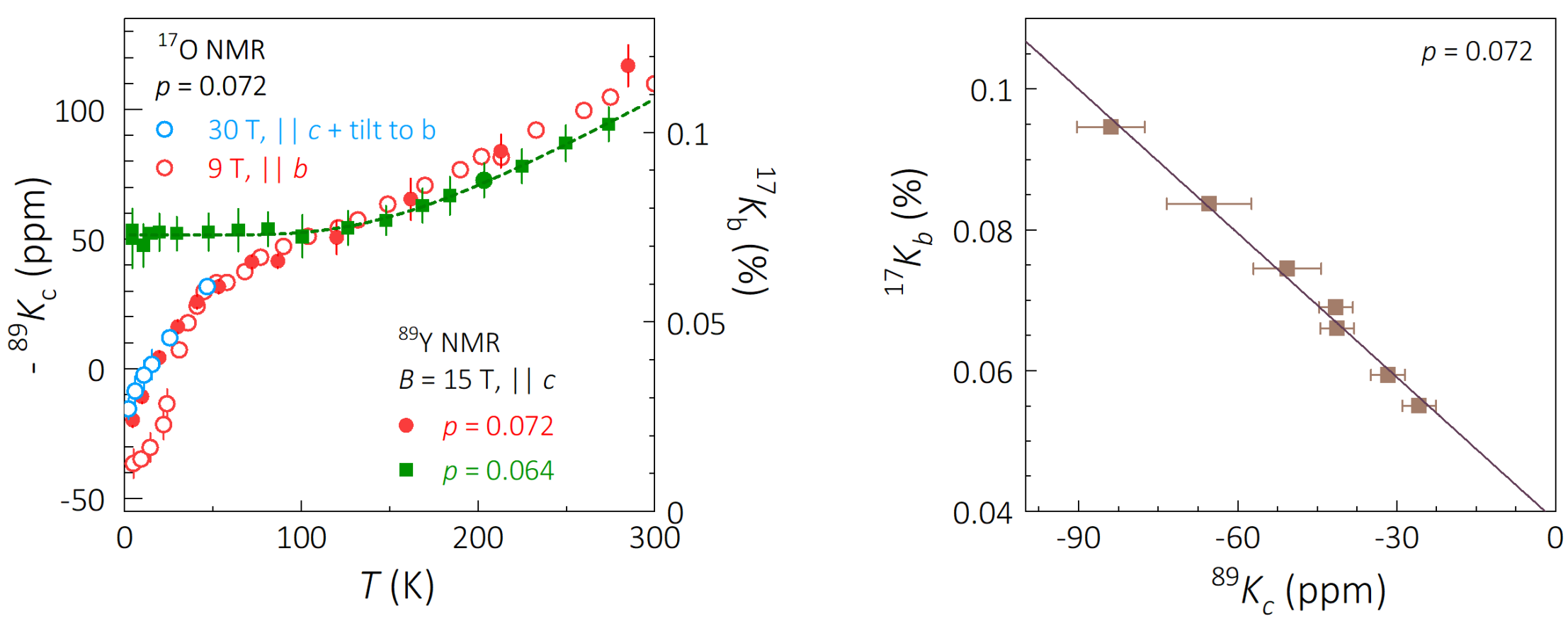

**Fig. S9. Scaling of $^{17}$O and $^{89}$Y Knight shifts.** (a). Temperature-dependent Knight shifts $^{89}K_c$ for $p$=0.064 and 0. 072 samples compared to $^{17}K_b$ for $p$=0. 072. (b) $^{89}K_c$ vs. $^{17}K_b$ in the normal state for the $p$=0.072 sample (values were interpolated as data points in (a) were not taken at the same temperatures). The data satisfies the relationship of $^{17}K_b$ = -681.34×$^{89}K_c$ + 0.03856%. Using this scaling relationship, we have rescaled $^{89}K_c$ in equivalent $^{17}K_b$ values, as shown in (a).

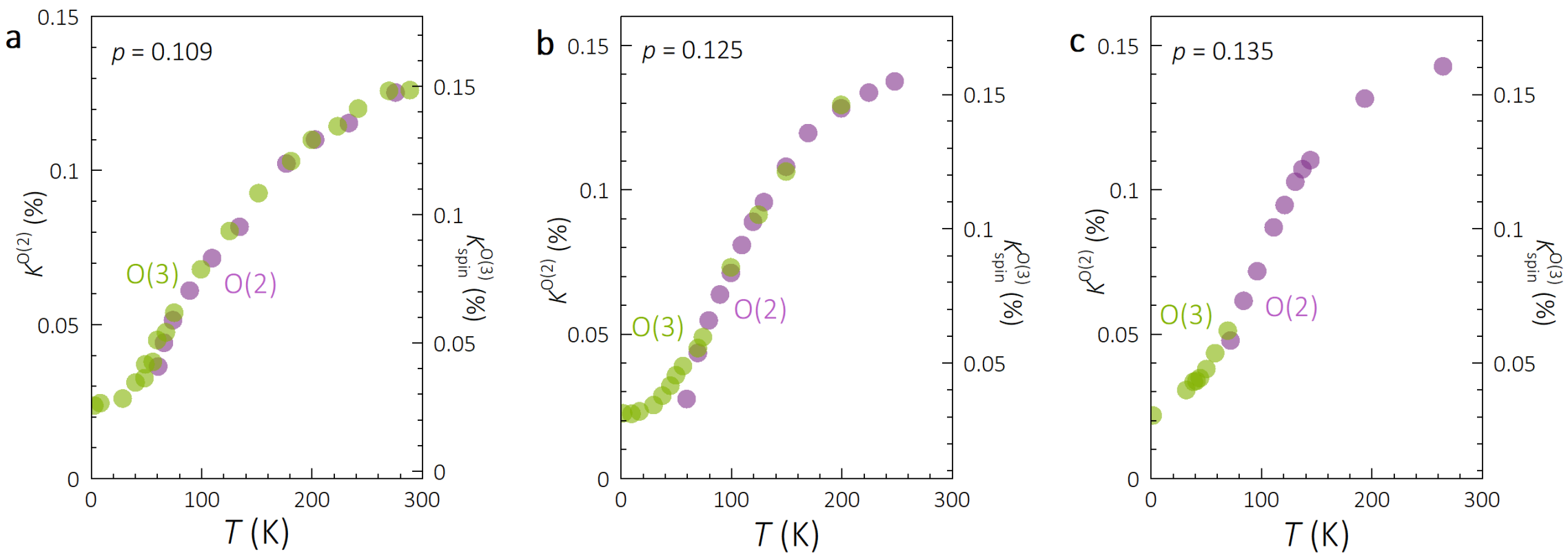

**Fig. S10. Scaling of O(2) and O(3) Knight shifts.** Comparison of $K$ for O(2) sites (measured with $B||b$) and $K_{spin}$ for O(3) sites (measured with $B$ tilted by 16°off the $c$ axis and $K_{orb}$ is subtracted) for $p$=0.109 (a), 0.125 (b) and 0.135 (c). Details about $K_{orb}$ for O(3) sites can be found in Ref. (9). For $p$=0.109, we find $K^{O(2)}(H||b)$= 0.093×$K_{spin}{}^{O(3)}$ + 0.002%. For $p$=0.125 and 0.135, $K^{O(2)}(H||b)$ = 0.093×$K_{spin}{}^{O(3)}$ – 0.007%. These relations give us $^{17}K_{orb}$ values for $B||b$ that are consistent with $^{17}K_{orb}$ = 0.006% (51).

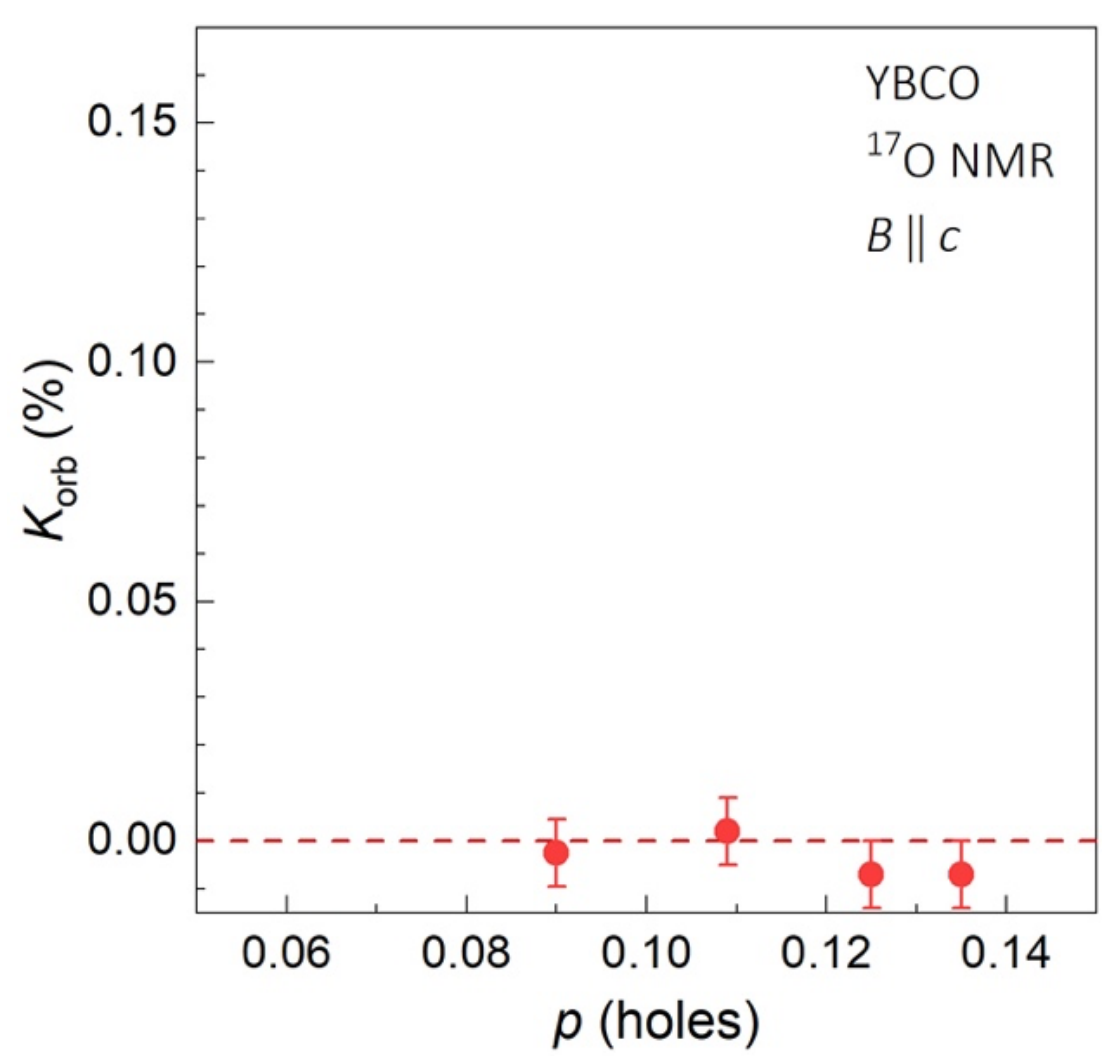

**Fig. S11. Orbital contribution to the $^{17}$O Knight shift (O(2) sites, $B||b$).** The values (red dots) are obtained from the scaling of O(2) and O(3) Knight shifts (Fig. S4). The data is shown in the same vertical scale as Fig. 2, in order to facilitate comparison. Clearly, $K_{orb}$ << $K_{spin}$ at any $T$ and can therefore be neglected.